\documentstyle[12pt]{article}
\begin{document}
\centerline                          
{\large\bf Approximate Flavor Symmetry in Supersymmetric Model} 
\vspace{0.8cm}
\centerline{\large Zhijian Tao}
\begin{center}
{ China Center of Advanced Science and Technology, (World Laboratory)\\
Theory Division, Institute of High Energy Physics, Academia Sinica\\
Beijing 100039, China}
\end{center}
\vspace{.2truecm}     

\begin{abstract}
We investigate the maximal approximate flavor symmetry in the framework of 
generic minimal supersymmetric standard model. 
We consider the low energy effective theory of the flavor physics 
with all the possible operators included. 
Spontaneous flavor symmetry breaking leads to the approximate flavor 
symmetry in Yukawa sector and the supersymmetry breaking sector. 
Fermion mass and mixing 
hierachies are the results of the hierachy of the flavor symmetry breaking. 
It is found that in this theory it is possible to solve the flavor 
changing problems. Furthermore baryogenesis of the universe can be well 
described and neutron electric dipole moment is closely below it 
experimental bound by assuming approximate CP violating phase $\sim 10^{-2}$ 
and superpartner mass around $100$ GeV. 
\end{abstract}
\renewcommand{\thefootnote}{\arabic{footnote}} \setcounter{footnote}{0} 
\newpage
In the Standard Model (SM), understanding of the patterns of fermion mass
and mixing remains as one of the most challenging problem. The Yukawa
couplings, with too many parameters, can account for the observed fermion
mass and mixing. However it does not answer the question why there exists
the observed fermion mass and mixing hierarchy. CP violation phase in
Cabbibo-Kobayashi-Maskawa (CKM) matrix is also a free parameter. One has no
idea about how large the phase should be from any first principle. And in
the framework of the SM this phase is determined to be order of one in order
to explain the CP violation measured in K meson system. However in the
extensions of the SM if there are other sources of CP violation, the CKM
phase is not determined at all. While the SM provides no understanding of
the fermion mass and mixing problem, from some physical consideration based
on flavor symmetry in some beyond SM it may provides an opportunity to
understand better this problem \cite{wil}. It has been noticed that in the
SM, the gauge interaction respects a maximal flavor symmetry $%
G_f=U(3)_Q\times U(3)_U\times U(3)_D\times U(3)_L\times U(3)_E$, where Q and
L are SU(2) doublet quarks and leptons, U, D, E are singlet quarks and
leptons respectively. Yukawa coupling sectors, though, do not respect this
symmetry (otherwise all the fermion masses are zero), the symmetry can be
treated as an approximate symmetry because all of the fermion masses except
top quark mass are much less than the weak scale. Natively one expects the
fermion mass and mixing hierarchies can be reproduced if the flavor symmetry
is spontaneously broken in steps $U(3)_f\stackrel{V_3}{\to }U(2)_f\stackrel{%
V_2}{\to }U(1)_f\stackrel{V_1}{\to }$nothing, with $V_3>>V_2>>V_1$. Here $f$
represents all kind of the fermions. The Yukawa coupling must be generated
through high dimensional non-renormalizable operators when some scalar
fields get vacuum expectation values (VEV) to break the flavor symmetry. The
hierarchies of fermion mass and mixing are the result of the hierarchy of
the flavor symmetry breaking. Indeed this idea has been investigated
recently by using a parameterization approach \cite{ant,hal}. It can
reproduce the fermion mass and mixing hierarchies in a sense of order of
magnitude. It naturally suppresses the flavor-changing neutral current
(FCNC) in some degree in multi-Higgs doublet models. However it should be
noticed that since the third generation is the heaviest one and the top
quark mass is at the weak scale, $U(3)_f$ symmetry can be strongly broken
down to $U(2)_f$. So $U(2)_f$ symmetry is a better approximation to describe
the fermion mass and mixing problem. In this case the first two generation
belongs to an $U(2)_f$ doublet and the third generation belongs to a singlet
of $U(2)_f$. As only concerning on the first two generations, this two
flavor symmetry approaches are effectively same.

It is well known that in the Minimal Supersymmetric Standard Model (MSSM),
the fermion mass and mixing problem remains exactly the same way as that in
the SM, therefore the flavor symmetry approach is also used to address this
problem in MSSM. Moreover $U(3)_f$ or $U(2)_f$ symmetries provide the needed
super-GIM mechanism in scalar-quark sector to suppress FCNC in MSSM, i.e.
the first two generation of scalar-quarks must be highly degenerate in order
to avoid too large FCNC \cite{gro}. With an exact flavor symmetry the three
(or the first two) generation scalar-quarks are precisely degenerate.
Breaking the flavor symmetry will spoil the degeneracy. How large is the
non-degeneracy of the first two generation scalar-quarks is therefore
related to the flavor symmetry breaking pattern and the quark mass and
mixing values. Based on solely the symmetry ground without assuming further
anything on the underlying theory, one may write down all possible
non-renormalizable operators allowed by the flavor symmetry and the $%
SU(2)_L\times U(1)$ gauge symmetry. Some operators contribute to the Yukawa
couplings and some contribute to scalar-fermion mass and mass splitting.
Generally the lower dimensional operators dominate over the higher
dimensional operators. When fermion mass and mixing are generated due to the
spontaneous flavor symmetry breaking, scalar-quark mass breaking is produced
too. Assuming all the fermions transform under the same flavor group U(3) or
U(2) it is found that typically the ratio of the mass square differences
over the mass square themselves of all type of the first two generation
scalar quarks are around the ratio of the $s$ quark and $b$ quark masses 
\cite{pom}. This leads to a problematic contribution to the $\epsilon_K$
parameter of the Kaon physics \cite{nir}. To comply with the $\epsilon_K $
parameter, one has to have the following choices \cite{pom}, 1) To specify
some concrete underlying models, so that the lower dimensional operators
contributing to the squark mass splitting is not allowed\cite{pom}. 2) The
squark masses are much heavier than weak scale, say $\geq 5$TeV. 3) CP
violating phases are small, $\sim 10^{-2}$, in this case the restriction on
squark masses are relaxed, but still they are required to be larger than
about $500$GeV \cite{pom}. 4) Supersymmetry is assumed to be broken by a
hidden sector, the widely studied scenario is the gauge-mediated
supersymmetry breaking, so that all the supersymmetry breaking interaction
terms are real \cite{din}.

In this work we investigate the maximal global flavor symmetry group $G_f$
in MSSM \cite{rand}. If $G_f$ is gauged it is well known that gauge contribution in D
terms will cause large splitting to the squark masses, so it spoils the
super-GIM cancellation mechanism. The crucial point in our theory is that
each type of quarks corresponds to different flavor symmetry. For instance
the D quark is put in the fundamental representation of $U(3)_D$ or $U(2)_D$%
. So the D type quark mass and scalar-quark mass generation can be
independent in some degree from other type of quarks and scalar-quarks. We
find that in MSSM with $tan\beta \leq 1$, the first two generation D type
squark mass splitting can be further suppressed by about two order of
magnitudes than the typical splitting in a universal flavor symmetry group
case for all the fermions, as the breaking scale of $U(3)_D$ is one order of
magnitude smaller than other breaking scales of $U(3)_f$. In fact we think
it is a very attractive idea in MSSM to address the problems of FCNC and
fermion mass generation simultaneously based on flavor symmetry
consideration. This leads to a very interesting scenario which may account
for many interesting questions in MSSM. Now we come to discuss the scenario.

First we clarify some general assumptions we are going to follow in this
work, which are:

1) The global flavor symmetry is the maximal flavor group $G_f=U(3)_Q\times
U(3)_U \times U(3)_D\times U(3)_L\times U(3)_E$. However, for most of the
discussion in the following, the results are the same if we choose U(2)
symmetry instead of U(3).

2) All the coupling constants in the theory are order of one, we expect the
fermion mass and mixing patterns are the results of the hierarchies of the
flavor symmetry breaking scales.

3) Flavor symmetry is broken only by gauge singlet scalar fields getting
VEV. The flavor symmetry is spontaneously broken in steps $U(3)_f\stackrel{%
V_3}{\to }U(2)_f\stackrel{V_2}{\to }U(1)_f\stackrel{V_1}{\to }$ nothing,
with $V_3>>V_2>>V_1$. Here $f$ represents all kind of the fermions.

4) Higgs doublets $H_u$ and $H_d$ which finally couple to up and down quarks
respectively are the singlets of $G_f$.

5) We don't specify any model of the flavor physics as the underlying
theory, instead we work at low energy scale taking into account all the
possible non-renormalizable operators allowed by the symmetry $G_f$. Only
thing here we assume is that there is a high scale of the flavor physics $%
\Lambda_f$. Above this scale one has to work with an underlying theory.
Below this scale one has an effective theory with non-renormalizable
operators suppressed by powers of the large scale $\Lambda_f$. $\Lambda_f$
is a scale much higher than the weak scale since spontaneous breaking of
global flavor symmetry leads to familons, which constrain $V_f>10^{10}$GeV 
\cite{tur}

The lowest dimensional $G_f$ invariant operators responsible for fermion
mass generation are, 
\begin{equation}
W_Y=g_U{\bar Q}U\frac{\phi ^Q\phi ^U}{\Lambda _f^2}H_u+g_D{\bar Q}D\frac{%
\phi ^Q\phi ^D}{\Lambda _f^2}H_d+g_E{\bar L}E\frac{\phi ^L\phi ^E}{\Lambda
_f^2}H_d 
\end{equation}
where $\phi ^f$ are gauge singlet and $U(3)_f$ triplet. These are dimension
six operators. Other representations of $G_f$ can only contribute to the
fermion mass through higher dimensional operators. Therefore they are more
suppressed by some powers of $\Lambda _f$. The higher dimensional operators
are important only if the lower dimensional operators are not allowed, which
is not the case for the low energy effective theory with most general
operators. The couplings $g_U$, $g_D$ and $g_E$ are assumed approximately to
be one. The SM Yukawa couplings are generated when all the $U(3)_f$
symmetries are broken down to nothing. When $\Lambda _f$ is put at infinity,
an exact $G_f$ symmetry is recovered at all energy scales. For a finite $%
\Lambda _f$, $G_f$ will be spontaneously broken below certain symmetry
breaking scale. At low energy scale an approximate flavor symmetry $G_f$ is
resulted. $U(3)_f$ symmetry can be spontaneously broken down 
through VEV's of $\phi ^f$, $<\phi ^f>=(V_1^f,V_2^f,V_3^f)$. 
In fact to break the $U(3)_f$ symmetry at one triplet scalar is 
not enough. One needs either to consider higher than six dimension operators
with other representation scalars of $G_f$ or to introduce another two 
set of $%
U(3)_f$ triplet scalar fields ${\phi ^{\prime }}^f$ and ${\phi''}^f$. 
In this work we
consider the model with ${\phi'}^f$ and ${\phi''}^f$ . 
The VEV's ${\phi'}^f$ and ${\phi''}^f$  
are expressed as 
$<{\phi'}^f>=({V_1'}^f,{V_2'}^f,{V_3'}^f)$ and 
$<{\phi'}^f>=({V_1''}^f,{V_2''}^f,{V_3''}^f)$.

Denoting ${\epsilon _f}_i=\frac{V_i^f}{\Lambda _f}$, ${\epsilon
_f'}_i=\frac{{V_i'}^f}{\Lambda _f}$, 
and 
${\epsilon_f''}_i=\frac{{V_i''}_f}{\Lambda _f}$
here $i=1,2,3$ is the
generation index, the Yukawa couplings generated from the flavor symmetry
breaking are given as 
\begin{equation}
Y^U_{ij}\simeq({\epsilon_Q}_i+{\epsilon_Q'}_i+{\epsilon_Q''}_i)
({\epsilon_U}_j+{\epsilon_U'}_j+{\epsilon_U''}_j)
\end{equation}
and likewise for $Y^D$ and $Y^E$. We expect the fermion mass and mixing
patterns are due to the hierarchy of the flavor symmetry breaking, 
$V_3^f$, $%
{V_3'}^f$, ${V_3''}^f>>V_2^f$, ${V_2'}^f$, 
${V_2''}^f>>{V_1}^f, {V_1'}^f, {V_1''}^f$. So we
have ${\epsilon _3}_{U,D,E},{\epsilon _3'}_{U,D,E}, 
{\epsilon _3''}_{U,D,E}
>>{\epsilon _2}_{U,D,E},{\epsilon _2'}_{U,D,E}, 
{\epsilon _2''}_{U,D,E}>>{\epsilon _1}_{U,D,E}, {\epsilon _1'}_{U,D,E}, 
{\epsilon _1''}_{U,D,E}$.
At this step, in fact we obtain the approximate flavor symmetry structure of
the Yukawa matrix, which is first proposed in Ref.\cite{ant}. In principle
higher dimensional operators also contribute to the Yukawa matrix.
Especially since the top quark is heavy, higher order of ${\epsilon _Q}_3$, $%
{\epsilon _U}_3$ should not be neglected. However it is still reasonable to
believe this Yukawa matrix is correct in the sense of the order of
magnitude. At least we can expect these matrices are good approximation for
the first two generations. If $U(3)_f$ symmetry is really badly broken in
the nature, one may start from the flavor $U(2)_f$ symmetry instead of $%
U(3)_f$. As we stated above for both cases the following discussions are the
similar.

In MSSM with $tan\beta \simeq 1$, all the relations of the $\epsilon $
parameters obtained in Ref.\cite{hal} by fitting the fermion masses and
mixing are the exactly the same here. We will adopt the values of the $%
\epsilon $ parameters in the following investigation. Given all these
parameters we can discuss the squark mass splitting and its implications.
Supersymmetry breaking terms are the most general ones allowed by the flavor
symmetry and the gauge symmetry of the theory. Before the flavor symmetry
breaking all the fermions are massless and various generation of
scalar-fermions are degenerate. Scalar-fermion mass terms are dimension two
operators at this step. Flavor symmetry must be broken to give fermion
masses, as the result the scalar-fermion masses split. The splitting of the
scalar-fermion masses are induced by the higher dimensional operators as
flavor symmetry is broken. The lowest dimensional operators have the form
like $m_{\tilde f}^2\tilde f^i\tilde f_j\frac{\phi _f^i{\phi _f}_j}{\Lambda
_f^2}$, etc. Hence the splitting are related to the $\epsilon $ and $%
\epsilon', \epsilon''$ parameters as, 
\begin{equation}
\frac{m_{\tilde f_i}^2-m_{\tilde f_j}^2}{m_{\tilde f_i}^2+m_{\tilde f_j}^2}%
\simeq (\epsilon _{f_i}+{\epsilon'}_{f_i}+{\epsilon''}_{f_i})^2
{}{}{}{}{}{}~~(i>j) 
\end{equation}
Hereafter we will only use $\epsilon $ parameter to represent the
contributions of all $\epsilon $, $\epsilon ^{\prime }$ and 
$\epsilon''$. This
corresponds to a redefining of $\epsilon $ as 
$\epsilon +\epsilon'+\epsilon''$%
. From this relation one sees that the scalar-fermion mass differences
between the third and the first two generations can be as large as the
scalar-fermion masses themselves because the third generation is heavy,
while the splitting between the first two generations are suppressed by $%
\epsilon _{f_2}^2$. The values of these parameters are estimated by fitting
quark masses and mixing angles as \cite{hal}

$$
\epsilon _{Q_1}/\epsilon _{Q_2}\simeq 0.2, \hspace{ 0.5cm}\epsilon
_{Q_2}/\epsilon _{Q_3}\simeq 0.04\hspace{0.5cm}\epsilon _{Q_1}/\epsilon
_{Q_3}\simeq 0.008%
$$
$$
\epsilon _{U_1}\epsilon _{Q_3}\simeq 0.04\frac V{<H_u>}, \hspace{0.5cm}\epsilon _{U_2}\epsilon _{Q_3}\simeq 0.2\frac V{<H_u>}%
$$
\begin{equation}
\epsilon _{D_1}\epsilon _{Q_3}\simeq 0.006\frac V{<H_d>},\hspace{0.5cm}\epsilon _{D_2}\epsilon _{Q_3}\simeq 0.025\frac V{<H_d>},\hspace{0.5cm}\epsilon _{D_3}\epsilon _{Q_3}\simeq 0.03\frac V{<H_d>} 
\end{equation}
here $V=175$GeV, $<H_u>$ and $<H_d>$ are VEVs of $H_u$ and $H_d$ fields. If $%
\tan \beta =1$ we go back to the result of Ref\cite{hal}. The fact that $%
\epsilon _{D_2}$ is much smaller than $\epsilon _{U_2}$ is because $c$ quark
is much heavier than $s$ quark. While a small $\epsilon _{D_2}$ implies a
small mass splitting for the first two generation scalar-down quarks $\delta
m_{\tilde D}^2$. This is very different from the model where all the
fermions are put in the same flavor symmetry group. Generally here in our
model $\delta m_{\tilde D}^2$ is about two order of magnitude smaller if $%
\tan \beta \simeq 1$. Due to the smallness of $\delta m_{\tilde D}^2$ we
expect the contribution to $\epsilon_K$ in the theory is not too large to
cause any problem.

Now we come to discuss the most interesting implications of our theory in
particle phenomenology

{\it Supersymmetric $\epsilon _K$ problem} \hspace{0.5cm} In SUSY models there
are additional CP violation sources which can contribute to K meson system
CP violation. The contributions to the CP violating $\epsilon _K$ parameter
of K meson system is dominated by diagrams involving Q, D squarks and gluino
in the same loop. For $m_{\tilde g}\simeq m_{\tilde Q}\simeq m_{\tilde
D}\simeq \tilde m$, here $m_{\tilde g}$ is gluino mass, and considering only
the contribution from the first two generation squark families, one gets 
\cite{nir,hag} 
\begin{equation}
\epsilon _K=\frac{5\alpha _3^2}{162\sqrt{2}}\frac{f_K^2m_K}{\tilde m^2\Delta
m_K}\left[\left(\frac{m_K}{m_s+m_d}\right)^2+\frac 3{25}\right] Im\left\{ 
\frac{V_{11}^Q\delta m_{\tilde Q}^2{V_{21}^Q}^{*}}{m_{\tilde Q}^2}\frac{%
V_{11}^D\delta m_{\tilde D}^2{V_{21}^D}^{*}}{m_{\tilde D}^2}\right\} 
\end{equation}
where $\delta m_{\tilde Q,\tilde D}^2$ are the mass square differences
between the first and second families of squarks. Flavor mixing matrices $%
V^{Q,D}$ are the coupling matrices in the vertex $\tilde g-\tilde Q-Q$ and $%
\tilde g-\tilde D-D$ where both quark and squark matrices are diagonal.
Using the experimental value of $\epsilon _K$, one gets 
\begin{equation}
(\frac{100{\rm {GeV}}}{\tilde m})^2\left [\frac{V_{11}^Q\delta m_{\tilde Q}^2%
{V_{21}^Q}^{*}}{m_{\tilde{Q}}^2}\frac{V_{11}^D\delta m_{\tilde D}^2{V_{21}^D}%
^{*}}{m_{\tilde{D}}^2}\right]\sin \phi \leq 0.5\times 10^{-8} 
\end{equation}
where $\phi =arg(V_{11}^Q{V_{21}^Q}^{*}V_{11}^D{V_{21}^D}^{*})$. In our
theory $\delta m_{\tilde{Q},\tilde{D}}^2$ is estimated from Eq.\cite{hal}
and $V^{Q,D}$ can also be estimated as the following 
\begin{equation}
V_{11}^{Q,D}\simeq 1,{}{}{}V_{21}^Q\simeq \frac{\epsilon _{Q_1}}{\epsilon
_{Q_2}}\simeq 0.2,{}{}{}V_{21}^D\simeq \frac{\epsilon _{D_1}}{\epsilon _{D_2}%
}\simeq 0.24 
\end{equation}
Therefore we obtain the following constraint 
\begin{equation}
(\frac{100{\rm {GeV}}}{\tilde m})^2(\frac V{<H_d>})^2\sin \phi \leq 0.1 
\end{equation}
This constraint can be easily satisfied with $\tan \beta \simeq 1$ either
requiring $\phi \leq 0.1$ or $\tilde m\geq 300$GeV. It is easy to check that
the contribution involving the third family squarks via (13) and (23) mixing
elements is comparable to that from the first two families.

{\it Neutron electric dipole moment $d_n$} \hspace{0.5cm} In SUSY model $d_n$
is dominated by the one loop gluino diagrams \cite{buc} 
\begin{equation}
d_n\simeq M_d\frac{e\alpha _3}{18\pi \tilde m^4}(|Am_{\tilde g}|\sin \phi
_A+\tan \beta |\mu m_{\tilde g}|\sin \phi _B) 
\end{equation}
here all the parameters have the conventional definitions in SUSY models. If
one only considers the first generation squark contribution. $M_d$ is the $d$
quark mass. Certainly the second and the third families of squarks may also
contribute to $d_n$ through the mixing matrices $V^{Q,D,U}$. Their
contributions to $M_d$ can be expressed as $V_{1i}^QV_{1i}^Dm_i$ (i=s,b) or $%
V_{1i}^QV_{1i}^Um_i$ (i=c,t). Using values of $V^{Q,D,U}$ and the quark
masses we find that the contribution from the second and third families of
squarks do not exceed the contribution from the first generation. Hence the
supersymmetric $d_n$ problem remains the same in our theory. In order to
avoid too large $d_n$, as usual one has to either require $\tilde m$ as
large as order of TeV, or CP violating phase as small as $10^{-2}$.

{\it Baryogenesis of the universe} \hspace {0.5cm} It is widely realized that
in order to generate the observed baryon asymmetry of the universe, by using
electroweak baryon number violation one needs some new physics beyond the SM
at the weak scale. The reason is that in the SM CP violation effect is not
strong enough and the electroweak phase transition is at best weakly first
order \cite{coh}. SUSY is the most promising extension of the SM in this
respect as it possibly provides a stronger electroweak phase transition and
some new sources of CP violation.

In MSSM the electroweak phase transition can be sufficiently strong first
order due to the large coupling of a light stop to the Higgs boson. And it
is demonstrated that the phase transition is strong enough whenever \cite
{car} 
\begin{equation}
m_h\stackrel{<}{\sim} 80{\rm {GeV},{}{} m_A\stackrel{>}{\sim} 200{GeV}%
,{}{}\tan\beta\stackrel{<}{\sim} 2.5,{}{} m_{\tilde{t}_R}\leq 175{GeV},{}{} 
\tilde{A}_t\simeq 0 }{} 
\end{equation}
here $m_h$ is the mass of the lightest Higgs boson, $m_A$ is that of the
pseudoscalar Higgs boson, and $\tilde{A_t}$ is the effective $\tilde{t}_L- 
\tilde{t}_R$ mixing parameter. These limits are slightly relaxed if one
includes two-loop QCD effects \cite{esp}.

Recent investigation shows that with MSSM parameters chosen in above region,
the new sources of CP violation in squark sector or Higgs sector may
generate the required amount of baryon asymmetry of the universe during the
electroweak phase transition \cite{wor}. Here we would like to point out
that all the conditions for the required baryogenesis can be satisfied in
our theory, while at the same time our theory provides a solution to $%
\epsilon_K$ and $d_n$ problems. To be more specific we present a model as an
example in the following.

First of all we assume $\tilde m\simeq 100{\rm {GeV}}$, so that new physics
effect from SUSY appears at weak scale. In order to evade the $d_n$ problem
we have to choose the CP violating phases around $10^{-2}$. This can be
naturally achieved by assuming that CP is an approximate symmetry in the
model. If so $\epsilon _K$ cannot be accounted for by CKM CP violating
phase. However the $\epsilon _K$ is possibly dominantly contributed by the
SUSY diagrams. In fact with $\tan \beta \simeq 3$ and $\tilde m\simeq 100%
{\rm {GeV}}$, $\phi \simeq 10^{-2}$ Eq.\cite{pom} gives the needed
contribution to $\epsilon _K$. With the small value of $\tan\beta $ and
given the squark and gluino masses at weak scale the required conditions of
baryogenesis can be filled. Form now on we discuss the experimental
consequences of the model with parameters chosen as above. As we have seen
this model may solve the supersymmetric $\epsilon _K$ and $d_n$ problems and
possibly provide an answer to the generation of the baryon asymmetry of the
universe.

Direct CP violating effects in the K meson decays will be contributed both
from CKM source and new sources in supersymmetric breaking sectors. CKM
contribution to $|\epsilon_K^{\prime}/\epsilon_K|$ will be order of $10^{-5}$
since the CKM phase is order of $10^{-2}$. However the contribution from
SUSY breaking terms is quite model dependent. It can be as large as $10^{-3}$
\cite{gro,nir,bab}. B factories are going to test the CKM mechanism of CP
violation. If the CP violating phase is as small as $10^{-2}$ in CKM matrix,
its CP violating effect is too small to be observed in B factory \cite{nir1}.

For $B^0-\bar B^0$, $D^0-\bar D^0$ and $K^0-\bar K^0$ mixing we estimate the
SUSY contributions to the $B$, $D$ and $K$ meson mass differences in our
model and find that their contributions are well below the experimental
values. For $B$ and $K$ systems the contributions from SUSY are smaller than
that from the SM. And for $D$ meson system SUSY contribution is close to the
contribution from the short distance SM box diagram. So we see that this
model is quite weakly constrained from the measurement of $B^0-\bar B^0$, $%
D^0-\bar D^0$ and $K^0-\bar K^0$ mixing. And it does not predict an
observable mixing effect in $D$ meson system in the future experiments. In
other words if the mixing effects of $D$ meson system is found in the future
experiments, it means this mixing effect is either from SM long distance
contributions or from some new physics effects beyond this model.

While neutron electric dipole moment $d_n$ is close to its experimental
bound in this model, we find the electron electric dipole moment can be
close to its lower experimental bound too. A recent work computes the
general contribution of one-loop electric dipole moment of electron in the
framework of MSSM \cite{nat}. It includes the full set of neutralino states
in the loop. Using the result of this work we find that with $\tilde{m}%
\simeq 100$GeV, $\phi\simeq 10^{-2}$ electron electric dipole moment is
estimated about $10^{-26}-10^{-27}$e-cm, which is close to its current
experimental bound. An improvement of the measurement of neutron and
electron dipole moments will possibly discover new physics effects studied
in this model.

Finally we consider the FCNC in lepton sector. The most interesting
processes are the lepton decays $\mu\to e\gamma$ and $\tau\to e(\mu)\gamma$.
These processes happen at one loop level where the sleptons and gauginos are
in the loops. Lepton family number is violated through the non-diagonal
elements of the slepton mass matrices or the left- and right-handed slepton
mixing matrices. The general formula for the decay rates of these processes
have been derived\cite{bor,his}. We use these formula to estimate these
decay rates in our model. As an reasonable estimation we choose all the new
mass parameters at $100$GeV, we find that the dominant contribution is from
the left-right mixing diagrams. This was pointed out in Ref.\cite{his},
though there a different framework is discussed. We estimate the decay rates
of $l_i\to l_j\gamma$ as, where $i,j$ are family index and $i>j$, 
\begin{equation}
\Gamma(\mu_i\to\mu_j\gamma)\simeq 2\times 
10^{-19}m_{l_i}^5{V^{L,E}_{ij}}^2{\rm GeV}^{-4} 
\end{equation}
Where $V^{L,E}$ are mixing matrice in the coupling vertex of
lepton-slepton- photino, $\tilde{\gamma}- l(E)-\tilde{L}(\tilde{E})$. In
terms of $\epsilon$ parametrs

\begin{equation}
V_{ij}^{L,E}\simeq \epsilon _i^{L,E}\epsilon _j^{L,E}
\end{equation}
These mixing matrices are not determined by solely fitting lepton masses.
Nevertheless if we assume $\epsilon _i^E\simeq \epsilon _j^L$ these
parameters can be approximated as 
\begin{equation}
\left( {\epsilon _1^{L,E}}\right)^2\simeq 3\times 10^{-6}\frac
V{<H_d>},~~~\left({\epsilon _2^{L,E}}\right)^2\simeq 6\times 10^{-4}\frac
V{<H_d>},~~~\left( {\epsilon _1^{L,E}}\right) ^2\simeq 10^{-3}\frac V{<H_d>}
\end{equation}
Then we predict the branching ratios as $Br(\mu \to e\gamma )\simeq 2\times
10^{-13}$ and $Br(\tau \to \mu \gamma )\simeq 10^{-11}$. All these decay
rates are too low to violate current experimental limits and it is difficult
to observe them in the near future experiments.

In summary we study the maximal flavor symmetry in the framework of MSSM. We
consider the low energy effective theory of the flavor physics with all the
possible operators included. In this theory we find the supersymmetric $%
\epsilon_K$ problem is solved and $d_n$ problem remains the same as in MSSM.
To fix both problems and to generate the required baryon number asymmetry of
the universe, we propose a model where all the SUSY particles are as heavy
as about $100$GeV and the CP violation phases are about $10^{-2}$. This
model is natural because the superpartners are lighter than TeV and the
small CP violation phase is protected by the CP symmetry. This model
predicts the electron electric dipole close to its current experimental
bound. Therefore we expect that about one order of magnitude improvement on $%
d_n$ and $d_e$ measurements may discover these new physics effects. As the
superpartners masses are around $100$GeV in the model, the next generation
collider experiments like LHC, Tevatron will certainly produce a lot of
numbers of these particles.

This work is supported by the National Science Foundation of China (NSFC).

\end{document}